\documentclass[aps,prc,groupedaddress,showkeys,showpacs,manuscript]{revtex4}
\usepackage{amssymb}

\usepackage{graphicx}

\begin{document}

\title{Transport model study of nuclear stopping in heavy ion collisions over an energy range from $0.09A$ GeV to $160A$ GeV}

\author {Ying Yuan,$\, ^{1,3}$\footnote{E-mail address: wawayubao@sina.com}
Qingfeng Li,$\, ^{2}$\footnote{E-mail address: liqf@hutc.zj.cn} Zhuxia Li,$\, ^{3}$\footnote{E-mail address: lizwux@ciae.ac.cn} and Fu-Hu Liu$\, ^{1}$\footnote{E-mail address: fuhuliu@163.com}}
\address{
1) Institute of Theoretical Physics, Shanxi University, Taiyuan, Shanxi 030006, China\\
2) School of Science, Huzhou Teachers College, Huzhou, Zhejiang 313000, China\\
3) China Institute of Atomic Energy, P. O. Box 275 (18), Beijing
102413, China }


\begin{abstract}
Nuclear stopping in the heavy ion collisions over a beam energy
range from SIS, AGS up to SPS is studied in the framework of the modified UrQMD transport
model, in which mean field potentials of both formed and
``pre-formed'' hadrons (from string fragmentation) and medium modified nucleon-nucleon elastic cross sections are considered. It
is found that the nuclear stopping is influenced by both the
stiffness of the equation of state and the medium modifications of
nucleon-nucleon cross sections at SIS energies. At the high SPS energies, the two-bump structure is shown in the experimental
rapidity distribution of free protons, which can be understood with
the consideration of the ``pre-formed'' hadron potentials.
\end{abstract}



\maketitle

{\section{Introduction}}

Since 1980's the heavy ion collisions (HICs) in terrestrial
laboratories have been becoming an important way to investigate
properties of hot and dense nuclear matter
\cite{Randrup1,Li2,Li3,Choi4,Bjorken5,Li6,Petersen7,Danielewicz8}.
In particular, the study of transport phenomena in nuclear reactions
is of major importance in the understanding of many fundamental
properties \cite{Escano9}. And, more interest was focused on
extracting the equation of state (EoS) of nuclear matter from the
comparison of microscopic transport models with experimental
measurements. Recently, the effect of medium modifications on
two-body collisions is received more and more attention.

In one of the attempts to obtain information about the EoS from
heavy ion data \cite{Danielewicz10}, it is made clear that progress
on this topic requires improved understanding of the momentum
dependence of mean fields generated in HICs as well as an extensive
modification according to experimental information on the degree of
stopping achieved \cite{Reisdorf11}. An optimal condition for nuclear matter compressed to form a dense medium is that the two
colliding heavy ions are fully stopped by each other during the
process of interaction, before the system starts to expand
\cite{Andronic12}. Information on the stopping can be obtained by
studying the rapidity distributions of fragments or free nucleons in
the transverse and longitudinal directions. In \cite{Reisdorf11},
the ratio of the widths of the transverse to the longitudinal
rapidity distributions was proposed as an indicator of the stopping degree.

The main purpose of this work is to extract the information of
nuclear stopping by the comparison of the rapidity distributions of protons and other stopping related observable from a transport-model simulation with data. Meanwhile,
medium modifications on interactions of particles in the dense
matter can be detected as well. This goal can be achieved by
studying the excitation function of the stopping from Au+Au
collisions at SIS energies and the rapidity distribution of free protons
from Au+Au/Pb+Pb collision at AGS and SPS energies, respectively,
within a transport model ---The Ultra-relativistic Quantum
Molecular Dynamics (UrQMD) model. The advantages of this method are:
(1) to directly compare existing data in each energy region, and (2)
to minimize the uncertainties coming from initial conditions and
final freeze-outs when more models are adopted.

{\section{UrQMD transport model}}

The UrQMD model is a microscopic many-body transport approach and
can be applied to study pp, pA and AA interactions over an
energy range from SIS to RHIC. This transport model is based on the covariant
propagation of color strings, constituent quarks and diquarks (as
string ends) accompanied by mesonic and baryonic degree of freedom
\cite{Petersen13}. In present model, the subhadronic degrees of
freedom enter via the introduction of a formation time for hadrons
produced in the fragmentation of strings
\cite{Andersson14,Nilsson15,Sjostrand16}, which are dominant at the
early stage of HICs with high SPS and RHIC energies. While at SIS
and AGS energies, the new particles are produced from the decay of
resonances. During the hadronic transport, it is known that two
ingredients should be taken into account with care if a better
comparison with data is needed: mean-field potential and two-body
scattering cross section of particles (e.g., Ref. \cite{Zhang17}).

\subsection{The mean-field treatments}

The UrQMD model is based on parallel principles as the quantum
molecular dynamics (QMD) model: hadrons are represented by Gaussian
wave packets in phase space and the phase space of hadron \emph{i}
is propagated according to Hamilton's equation of motion
\cite{Bass18},
\begin{equation}
\dot{\vec{r_{i}}}=\frac{\partial{H}}{\partial{\vec{p_{i}}}},
\hspace{1cm}
\dot{\vec{p_{i}}}=-\frac{\partial{H}}{\partial{\vec{r_{i}}}}.
\label{eq:1}
\end{equation}
Here, $\vec{r}$ and $\vec{p}$ are the coordinate and momentum of the
hadron \emph{i}, respectively. The Hamiltonian $H$ consists of the
kinetic energy $T$ and the effective interaction potential energy
$U$,
\begin{equation}
H=T+U.  \label{eq:2}
\end{equation}
In the standard UrQMD model, the potential energy $U$ includes
the two-body and three-body Skyrme-, Yukawa-, Coulomb- and
Pauli-terms \cite{Bass18,Bleicher19},
\begin{equation}
U=U_{\rm sky}^{(2)}+U_{\rm sky}^{(3)}+U_{\rm Yuk}+U_{\rm Cou}+U_{\rm pau}.
\label{eq:3}
\end{equation}

For a better description of experimental data at SIS energies, more
potential terms have to be considered \cite{Li20}. In the modified version of UrQMD (based on the version $2.0$),
the following two terms are further added: (1)
the density dependent symmetry potential term $U_{\rm sym}$ and (2) the momentum-dependent term $U_{\rm md}$ \cite{Bass21}.
Both the potential terms
are very important for the dynamics of the intermediate-energy
neutron-rich HICs. In this work four parameter sets for EoS are
used for comparison: H-EoS, S-EoS, HM-EoS and SM-EoS, which can be
found in Ref. \cite{Li20}.

At higher beam energies (AGS and SPS energies), the Yukawa-, Pauli- and
symmetry-potentials of baryons become negligible, while the Skyrme-
and the momentum-dependent parts of potentials still influence the
whole dynamical process of HICs \cite{Li22}. At SPS energies, the
new production mechanism of particles (string fragmentation) plays
more and more important role, in which the formation time of hadrons
from the string fragmentation is determined by a ``yo-yo" mode
\cite{Bass18,Bleicher19}. During the formation time, the
``pre-formed" particles (string fragments that will be projected
onto hadron states later on) are usually treated to be
free-streaming, while reduced cross sections are only included for
leading hadrons. In previous calculations
\cite{Bass18,Bleicher19,Bratkovskaya23}, the interaction of the
newly produced ``pre-formed" particles is not taken into account.
Recently, the mean-field potentials for both formed and
``pre-formed" particles are considered for a better
understanding of HBT time-related puzzle \cite{Li24}. Meanwhile, in
Ref. \cite{Li24}, the rapidity distribution of net-protons from HICs at
the SPS energy $158A$ GeV is shown to have a two-bump structure with
the consideration of the ``pre-formed'' hadron potentials, which explains data fairly well. In this paper, more analyses about free protons at all SPS energies will be shown.

At AGS and SPS energies, the relativistic effect on the relative
distance and the relative momentum and a covariance-related reduced
factor used for the update of potentials \cite{Li22,Isse25} are
considered in calculations.

{\subsection{The in-medium nucleon-nucleon (NN) elastic cross sections}}

Besides the updates of the mean field part mentioned above, the influence of the medium modification on two-nucleon
cross sections at the intermediate energy region should be also considered. In the present work we consider medium modifications on nucleon-nucleon (NN) elastic cross sections in the modified UrQMD model. For the inelastic channels, we
still use the experimental free-space cross sections. It is
believed that this assumption has minor effect on our present
study at SIS energies. At present, three forms of in-medium NN elastic cross sections are considered, they are (1) $\sigma^{\rm free}$, the free nucleon-nucleon elastic cross
section. (2) ${\sigma_{1}}^{*}$, which is based on the extended QHD
theory and reads as \cite{Li26,Li27}
\begin{equation}
{\sigma_{1}}^{*}=F(u, \alpha, p)\sigma^{\rm free}, \label{eq:4}
\end{equation}
where the medium correction factor $F$ depends on the nuclear
reduced density $u={\rho_{i}}/{\rho_{0}}$, the isospin-asymmetry
$\alpha={(\rho_{n}-\rho_{p})}/{\rho_{i}}$, and the relative momentum
of two colliding nuclei. The $\rho_{i}$, $\rho_{n}$ and $\rho_{p}$
are the nuclear, neutron and proton densities, respectively. More
explicitly, the factor $F$ is \cite{Li26,Li27}
\begin{equation}
F(u, \alpha, p)={F_{u}}^{p}\cdot{F_{\alpha}}^{p} , \label{eq:5}
\end{equation}
where
\begin{equation}
\left\{%
\begin{array}{ll}
    {F_{u}}^{p}=1+{[\frac{2}{3}exp(-{u/0.54568})-\frac{2}{3}]}/{[1+{({p_{\rm NN}/p_{0}})}^{\kappa}]}, & \hbox{$p_{\rm NN}\leq$1 GeV/c;} \\
    {F_{\alpha}}^{p}=1+{[\tau_{ij}\eta({0.85/{(1+3.25u)}}){\alpha}]}/{[1+{({p_{\rm NN}/p_{0}})}^{\kappa}]}, & \hbox{$p_{\rm NN}\leq$1 GeV/c;} \\
    {F_{\alpha, u}}^{p}=1, & \hbox{$p_{\rm NN}>$1 GeV/c.}
    \label{eq:6}\\
\end{array}%
\right.
\end{equation}
Here $p_{\rm NN}$ is the relative momentum in the NN center-of-mass
system; $\tau_{ij}=-1$, $+1$, and $0$ in the case of $i=j=p$,
$i=j=n$, and $i\neq{j}$, respectively; $\eta$ is set to $-1$ for a nonrelativistic typed
splitting on the proton-proton and neutron-neutron elastic cross
sections in the isospin-asymmetric nuclear medium. The other
parameters $p_{0}$ and $\kappa$, which influence the slope of the
momentum dependence of the reduction factor $F_{u}$, are still with
somewhat uncertainty \cite{Li27}. In this work, we choose $p_{0}=0.5$
GeV$/c$ and $\kappa=6$ as an example. Employing this approach, it was found that the in-medium NN elastic cross sections were suppressed seriously at low relative momenta than at higher one depending on the medium density, which is similar to the Brueckner relativistic approach  \cite{Fuchs36, Gaitanos37}. (3) ${\sigma_{2}}^{*}$, as in Ref. \cite{Klakow28}, which reads as

\begin{equation}
{\sigma_{2}}^{*}=(1-\xi u){\sigma}^{\rm free}, \label{eq:7}
\end{equation}
where $\xi=0.5$ for $E_{\rm lab}<0.25A$ GeV in this work. It is easy to find that
the momentum constraint is not considered in ${\sigma_{2}}^{*}$. Further, the density
dependence of ${\sigma_{2}}^{*}$ is stronger than that of ${\sigma_{1}}^{*}$.

\begin{figure}
\includegraphics[angle=0,width=18.8cm]{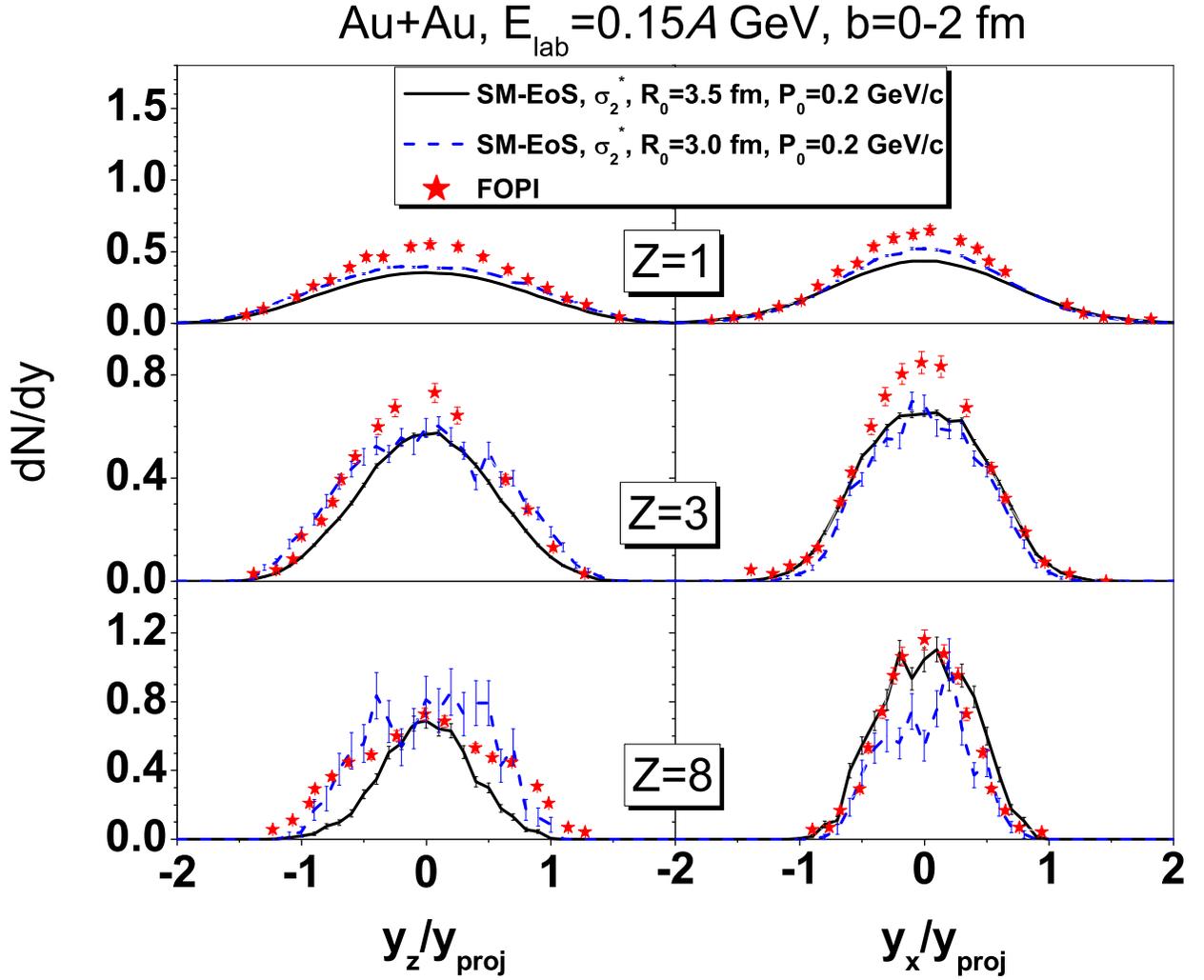}
\caption{Normalized rapidity distributions of fragments with proton
number Z=1, 3, and 8 (from top to bottom panels) in the longitudinal (left
panel) and transverse (right panel) directions for central Au+Au
collisions at $0.15A$ GeV. Two ($R_{0}$, $P_{0}$) parameter sets,
(3.5 fm, 0.2 GeV$/c$) and (3.0 fm, 0.2 GeV$/c$), are adopted in the
coalescence-model calculations, which are shown with lines. FOPI data
\cite{Andronic12} are shown by scattered stars.} \label{fig1}
\end{figure}

For calculations at SIS, a conventional phase-space coalescence
model \cite{Kruse29} is used to construct clusters, in which
nucleons with relative distances smaller than $R_{0}$ and relative
momenta smaller than $P_{0}$ are considered to belong to one
cluster. Fig.\ \ref{fig1} shows normalized rapidity distributions of
fragments with proton number Z=1, 3, and 8 (from top to bottom panels) in
the longitudinal (left panel) and transverse (right panel)
directions for central Au+Au collisions at $0.15A$ GeV. Two
($R_{0}$, $P_{0}$) parameter sets, (3.5 fm, 0.2 GeV$/c$) and (3.0
fm, 0.2 GeV$/c$), are adopted in the calculations. The results are shown
with lines and the FOPI data \cite{Andronic12} are shown by scattered stars. It seems that the parameter set (3.0 fm,
0.2 GeV$/c$) gives a better description of the FOPI data.
Therefore, this parameter set is used in the following calculations at
SIS energies in this work. While at AGS and SPS energies, the coalescence
model is not used as usual (partly because of the rich production of
new baryons) so that all nucleons at freeze-out are taken to be
free.

{\section{Nuclear stopping and the rapidity distributions}}

{\subsection{$vartl$ at SIS energies}}

As a measure of the nuclear stopping degree \cite{Li6}, the FOPI
Collaboration \cite{Reisdorf11} introduced a new observable
$vartl$ which was defined by the ratio of the variances of the transverse
to the longitudinal rapidity distributions of fragments. For central Au+Au
collisions, it is found that the rapidity distributions in the $x$
and $y$ directions are nearly the same, thus the transverse rapidity distributions are plotted approximately with the rapidity distributions in the $x$ direction. Numerically, the $vartl$ is defined as
\begin{equation}
{vartl}={\Gamma_{dN/dy_{x}}}/{\Gamma_{dN/dy_{z}}}, \label{eq:11}
\end{equation}
where $\Gamma_{dN/dy_{x}}$ ($\Gamma_{dN/dy_{z}}$) is the width of the rapidity
distribution of fragments in the $x$ ($z$) direction and reads as
\begin{equation}
\Gamma_{dN/dy_{x,z}}=\sqrt{\langle{y_{x,z}^{2}}\rangle}, \label{eq:12}
\end{equation}
\begin{equation}
\langle{y_{x,z}^{2}}\rangle=\frac{\sum{(y_{x,z}^{2}}{N_{y_{x,z}})}}{N_{\rm all}}. \label{eq:13}
\end{equation}
Here $N_{y_{x,z}}$ and $N_{\rm all}$ are yields of fragments in each $y_x$
(or $y_z$) rapidity bin and in the whole rapidity region, respectively. It is easy to understand that $vartl < 1$ stands for an incomplete stopping
or nuclear transparency, and $vartl >1$ for a strong transverse
expansion or collectivity. Obviously, $vartl =1$ when a full stopping occurs.

\begin{figure}
\includegraphics[angle=0,width=8.6cm]{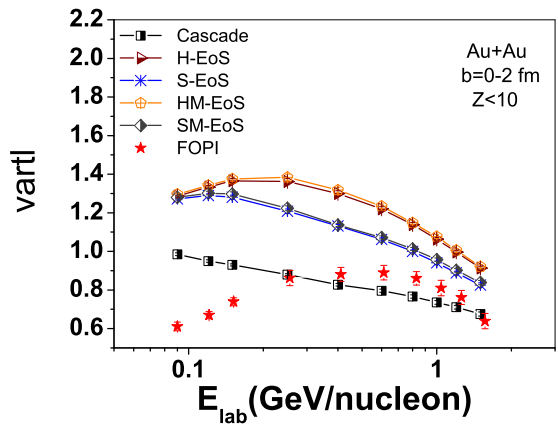}
\caption{Excitation function of $vartl$ for central Au+Au collisions at SIS energies. The FOPI data \cite{Reisdorf11}
are shown by stars while the UrQMD calculations with various
EoS are shown by lines with symbols.} \label{fig2}
\end{figure}

The excitation function of $vartl$ for central Au+Au collisions
is shown in Fig.\ \ref{fig2} within the beam energy region $0.09A-1.5A$ GeV . The FOPI
data \cite{Reisdorf11} are shown by stars while the
UrQMD calculations are shown by lines with
symbols. The $vartl$ value is calculated for fragments with the
proton number $Z<10$. In the calculations, results with the cascade
mode and with various EoS are shown. The free NN cross
sections are adopted in the calculations. It is seen that the $vartl$ value of the
cascade mode is always less than $1$ and decreases monotonously with the
increase of beam energy, which implies less and less stopping strength in
the system. At $E_{\rm lab} \sim 0.3A-1A$ GeV calculated values of
$vartl$ are smaller than data while it is larger than data at lower
beam energies. When the mean field is considered, the potentials reinforce the bound of nucleons and a stronger collectivity is shown
in the transverse direction. Among the calculations with EoS, softer
EoS gives a smaller $vartl$ value while the momentum dependent term
in the potential plays a negligible role. We also find that only a
soft EoS can not describe the excitation function of the FOPI data
without considering medium modifications of two-body collisions.
Next, based on the result with the SM-EoS, we will give a further
investigation of the effect of the medium modifications of NN
elastic cross sections on the $vartl$.

\begin{figure}
\includegraphics[angle=0,width=8.6cm]{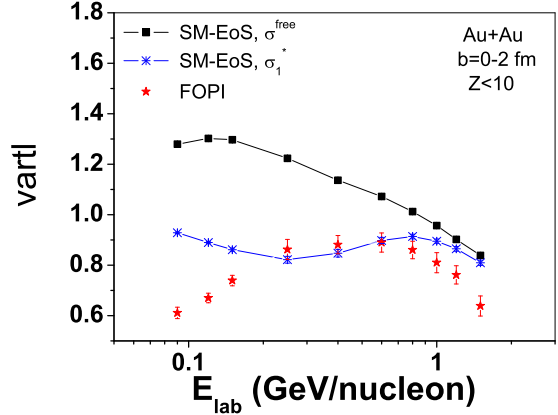}
\caption{Excitation function of $vartl$ with the medium modified NN
elastic cross section ${\sigma_{1}}^{*}$ as well as the free one
$\sigma^{\rm free}$. The SM-EoS is adopted in calculations. FOPI data \cite{Reisdorf11} are shown for comparison.}
\label{fig3}
\end{figure}

Fig.\ \ref{fig3} illustrates the calculated excitation function of $vartl$ with
the medium modified NN elastic cross section ${\sigma_{1}}^{*}$ as
well as the free one $\sigma^{\rm free}$. It is seen clearly that a
large reduction of cross sections at lower beam energies leads to
obvious transparency so that the calculated $vartl$ with
${\sigma_{1}}^{*}$ are largely decreased at low SIS energies. While
at high SIS energies the $vartl$ value is much less affected and
slightly higher than data. As mentioned in Eq.\ \ref{eq:6}, this
might be due to the fixed $p_{\rm NN}$ cut adopted. We would not modify
this just for fitting data since the medium modifications on
inelastic channels are still an open question. We just wish to
stress the importance of medium modifications of cross sections on
the nuclear stopping at moderate SIS energies.

\begin{figure}
\includegraphics[angle=0,width=8.6cm]{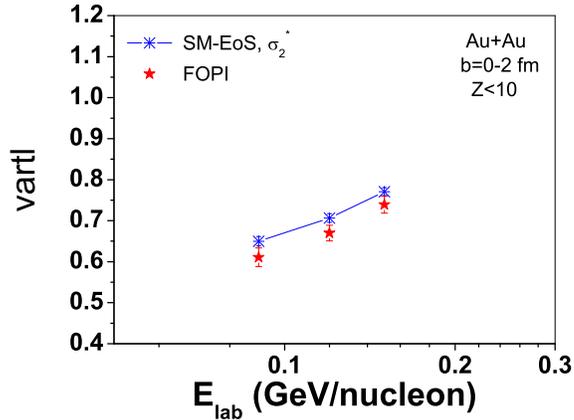}
\caption{Comparison of the FOPI data \cite{Reisdorf11} with calculations with
${\sigma_{2}}^{*}$ for $E_{\rm lab}<0.25A$ GeV. The SM-EoS is adopted in calculations.} \label{fig4}
\end{figure}

For $E_{\rm lab}<0.25A$ GeV, the results with ${\sigma_{1}}^{*}$ are still higher than data which implies that a
stronger reduction factor on the elastic cross sections is required.
Fig.\ \ref{fig4} further shows the calculation with
${\sigma_{2}}^{*}$ (with a stronger reduction factor on the NN
elastic cross section, as seen in Eq.\ \ref{eq:7}) for
$E_{\rm lab}<0.25A$ GeV. The comparison with data is fairly well and
same as done in Ref. \cite{Zhang17}.

{\subsection{Rapidity distribution at AGS and SPS energies}}

\begin{figure}
\includegraphics[angle=0,width=18.8cm]{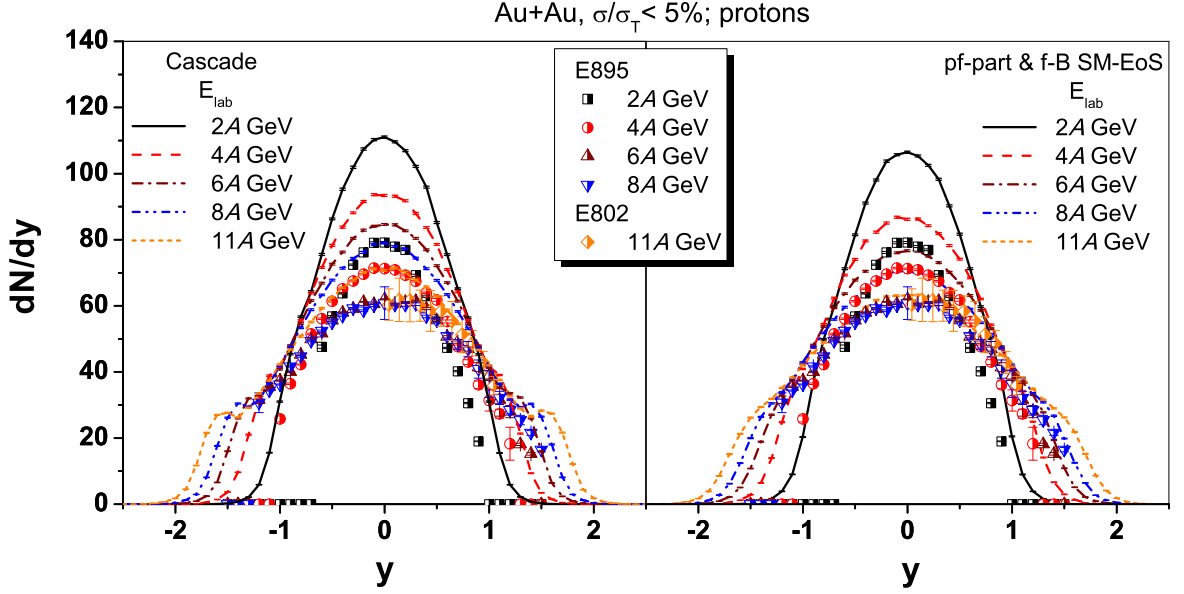}
\caption{Rapidity distributions of protons at AGS energies
$2A$, $4A$, $6A$, $8A$, and $11A$ GeV for central Au+Au
collisions. Calculations with cascade (left panel) and with potentials ``pf-part
\& f-B SM-EoS'' (right panel) are shown with lines. Experimental data of free
protons taken from E895 \cite{Klay30} and E802 \cite{Akiba31}
Collaborations are shown with scattered symbols.} \label{fig5}
\end{figure}

\begin{figure}
\includegraphics[angle=0,width=18.8cm]{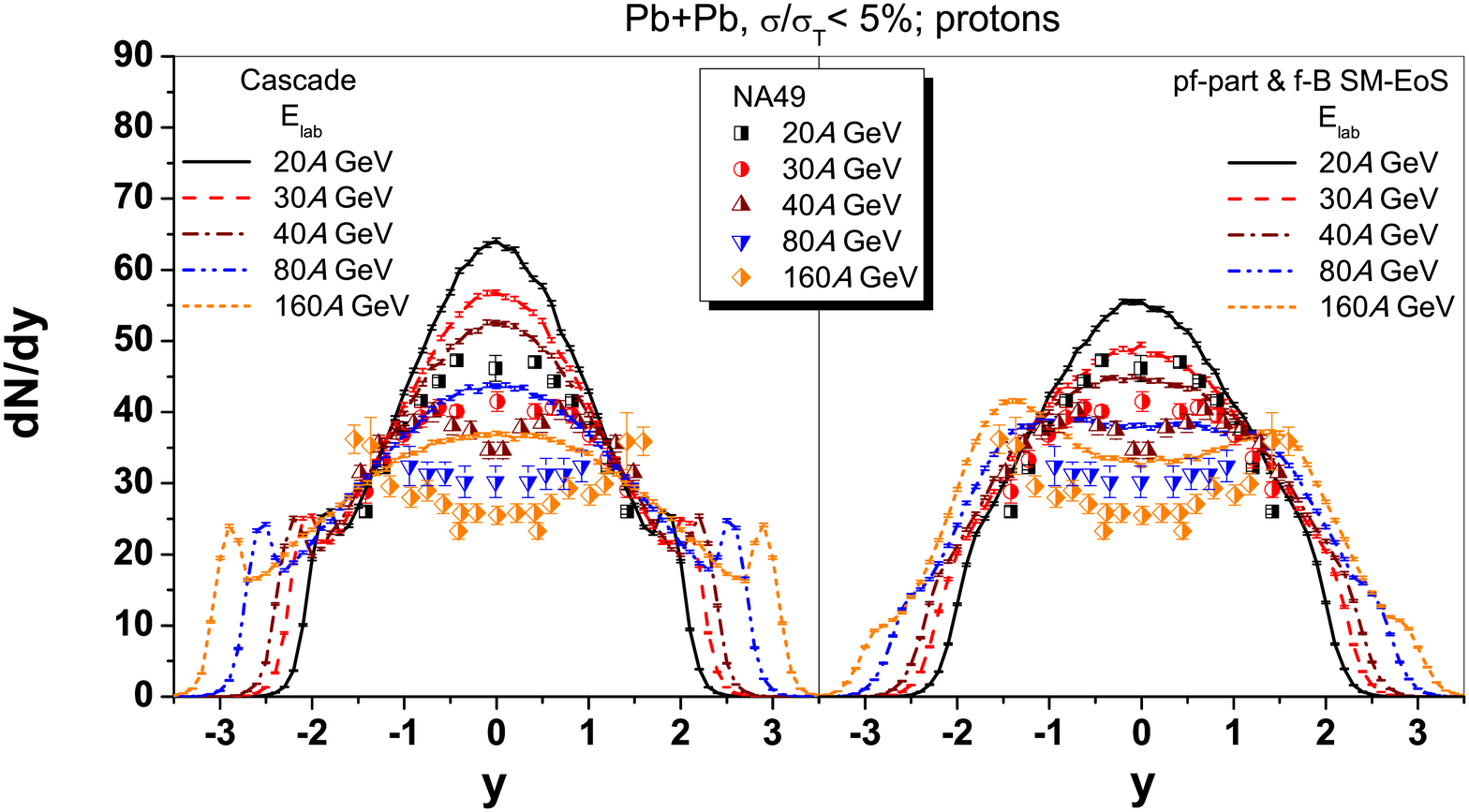}
\caption{Rapidity distributions of protons at SPS energies
$20A$, $30A$, $40A$, $80A$, and $160A$ GeV for central
Pb+Pb collisions. Calculations with cascade (left panel) and with potentials
``pf-part \& f-B SM-EoS'' (right panel) are shown by lines. Preliminary data
of free protons taken from NA49 \cite{Blume32,Strobele35}
Collaboration are shown by scattered symbols. } \label{fig6}
\end{figure}

At AGS and SPS energies, as the rapidity distribution of fragments in the transverse direction has not been provided by experiments, we study the nuclear stopping with the longitudinal rapidity distribution. Figs.\ \ref{fig5} and \ \ref{fig6} depict the rapidity distributions
of protons for central Au+Au collisions at AGS and for central Pb+Pb collisions at SPS ($<5\%$ of total cross section
$\sigma_{\rm T}$), respectively. The (preliminary) experimental data of free protons are
taken from \cite{Klay30,Akiba31,Blume32,Strobele35}. In the calculations,
besides a cascade mode shown in the left panel, we also show the
results with potentials of both formed and ``pre-formed" hadrons
(``pf-part \& f-B SM-EoS'') in the right panel. Cross sections used in the model
are not modified by the nuclear medium in this energy region. Since protons belonging to fragments are included in calculations of the rapidity
distribution, the calculation results of the proton number are somewhat larger than data, especially
at low beam energies as shown in Figs.\ \ref{fig5} and \ \ref{fig6} as well as in previous calculations \cite{Petersen13,Reisdorf33}. In Fig.\ \ref{fig5}, it is found that the shape of the rapidity distributions of measured protons changes from one peak at mid-rapidity with no shoulder to two shoulders when increasing beam energy from $2A$ GeV to $11A$ GeV. The cascade calculations always give a Gaussian-like distribution at $y<1.0$, while calculations with potentials are much closer
to data. With the increase of beam energy from AGS to SPS, the experimental rapidity distribution changes further to a plateau and finally to a two-bump structure.  Again, the calculations with cascade mode cannot describe the shape of the rapidity distribution of protons completely. The stronger repulsion at early stage introduced by potentials makes a wider rapidity distribution of protons in the longitudinal direction \cite{Li24}. The gap of two peaks becomes wider with the increase of beam energy. Especially, at $160A$ GeV the rapidity distribution of protons shows clearly two peaks at $y\sim1.5$. These features can be reasonably reproduced by the calculations with both the formed and ``pre-formed'' hadron potentials shown in the right panel of Fig.\ \ref{fig6}.

\begin{figure}
\includegraphics[angle=0,width=9.6cm]{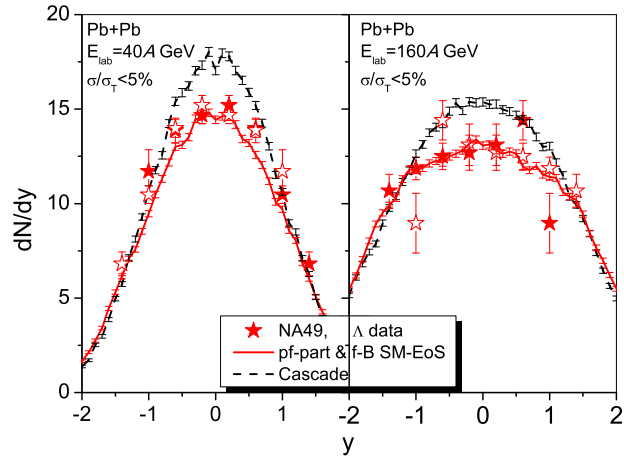}
\caption{Rapidity distributions of $\Lambda$s from central Pb+Pb
collisions at $40A$ GeV (left panel) and $160A$ GeV (right panel).
Calculations with and without potentials (lines) are compared to
the NA49 data \cite{Anticic34} (stars).  The open stars are data points reflected around mid-rapidity.} \label{fig7}
\end{figure}

We also calculate the rapidity distribution of emitted $\Lambda$s for central Pb+Pb collisions at $40A$ GeV and $160A$ GeV with and without formed and ``pre-formed'' hadron potentials as shown in Fig.\ \ref{fig7}.
Calculations with and without potentials (lines) are
compared to the NA49 data \cite{Anticic34} (stars). Same as data, the yields represent the sum $\Lambda+\Sigma^0$. It is seen
clearly that calculations with potentials are in good agreement with data at both beam energies, which is due to a larger transparency introduced by the strongly repulsive mean field at the early stage. As is known that at the AGS and SPS energies the yields
of hyperons are somewhat overestimated in the UrQMD cascade calculations with version less than $2.1$ \cite{Bass18,Bleicher19,Bratkovskaya23}, which is also shown in Fig.\ \ref{fig7}. In order to solve this problem, alternatively, starting from
the version $2.1$ (and the recently published v$2.3$), the UrQMD group considers additional high mass resonances that are
explicitly produced and propagated in s-channel processes with invariant masses up to $\sqrt{s}<3$ GeV\cite{Bratkovskaya23,Petersen:2009zi}. This treatment leads to lower yield of the strange particles so that a nice agreement with $\Lambda$ data from central Pb+Pb collisions at SPS energies was also shown in previous calculations \cite{Petersen13,Petersen:2009zi}. Therefore, it deserves much more investigations to deeply understand the effects of mean field potentials and the decay of high mass resonances on, e.g., particle production and collective flows, which are in progress.

{\section{Summary and Outlook}}

In summary, we have presented the excitation function of the nuclear stopping described by $vartl$ of light fragments for central Au+Au
reactions with beam energies from $0.09A$ GeV to $1.5A$ GeV and the
rapidity distribution of protons and $\Lambda$s for central
Au+Au/Pb+Pb reactions in the energy region $2A-160A$ GeV. The modified UrQMD
transport model (based on the version $2.0$) has been used in all calculations. Based on the model
we investigate the effects of both the mean-field potentials and
medium modifications of nucleon-nucleon elastic cross sections on
the nuclear stopping under the same initial and final freeze-out
conditions. It is found that the nuclear stopping is influenced by
both the stiffness of the equation of state and the medium
modifications of nucleon-nucleon elastic cross sections for reactions at SIS energies. And it reaches a well defined plateau of maximal stopping centered around $(0.5\pm0.3)A$ GeV with a fast drop on both sides. At AGS and SPS energies, the degree of nuclear stopping decreases continuously. In the high SPS energy region, as the transparency of matter is high, the two-bump structure is
shown in the experimental rapidity distribution of free protons in the longitudinal direction. Our calculations show that considering the potentials of both formed and ``pre-formed'' hadrons can improve the agreement between calculation results and data. But the form of the potentials is still simple and rough and further improvement is needed. The work on this aspect is underway.

{\section*{Acknowledgements}}

We acknowledge support by the Frankfurt Center for Scientific
Computing (CSC). This work is supported by: the Key Project of the
Ministry of Education of China under Grant No. 209053, the National
Natural Science Foundation of China under Grant Nos. 10675077, 10975095, 10675172, 10875031, 10905021, 10979023, the National Basic Research Program of China under Grant No. 2007CB209900, and the Natural Science Foundation of Zhejiang Province under grant No. Y6090210.

\end{document}